\documentclass[aps,pre,twocolumn,showpacs,floatfix,preprintnumbers,superscriptaddress,amsmath,amssymb]{revtex4-1}
\usepackage{amsmath}
\usepackage{amssymb}
\usepackage{graphicx}
\usepackage{epsfig}
\usepackage{dcolumn}
\usepackage{bm}
\usepackage{array}
\usepackage{xcolor}
\usepackage{physics}

\begin{document}
\title{Exact distribution of spacing ratios for random and localized states in quantum chaotic systems}
\author{S. Harshini Tekur}
\email{E-mail: harshini.t@gmail.com}
\affiliation{Indian Institute of Science Education and Research, 
Dr. Homi Bhabha Road, Pune 411 008, India}
\author{Santosh Kumar}
\email{E-mail: skumar.physics@gmail.com}
\affiliation{Department of Physics, Shiv Nadar University, Gautam Buddha Nagar, Uttar Pradesh 201314, India}
\author{M. S. Santhanam}
\email{E-mail: santh@iiserpune.ac.in}
\affiliation{Indian Institute of Science Education and Research, 
Dr. Homi Bhabha Road, Pune 411 008, India}

\begin{abstract}

Typical eigenstates of quantum systems, whose classical limit is chaotic,
are well approximated as random states. Corresponding eigenvalue spectra is modeled through
appropriate ensemble of random matrix theory.
However, a small subset of states violate this principle
and display eigenstate localization, a counter-intuitive feature known to 
arise due to purely quantum or semiclassical effects. In the spectrum of chaotic
systems, the localized and random states interact with one another and modifies
the spectral statistics. In this work, a $3 \times 3$ random matrix model is used to obtain
exact result for the ratio of spacing between a generic and localized state. 
We consider time-reversal-invariant as well as non-invariant scenarios. These
results agree with the spectra computed from realistic physical systems that display
localized eigenmodes.

\end{abstract}

\maketitle

\section{INTRODUCTION}

The generic eigenstates of quantum systems, whose classical limit is chaotic,
display uniform probability density, except for feature-less 
fluctuations \cite{uprob,sch}.
Physically, this reflects the underlying irregular dynamics of a typical classical
trajectory in agreement with the correspondence principle. 
It is also well-known that the eigenvalues of consecutive generic eigenstates
tend to repel one another in accordance with the Bohigas-Giannoni-Schmidt
conjecture \cite{bgs} and is modeled through random matrix theory (RMT) \cite{mehta2004,forrester2010}. 
This level repulsion property, encoded in the Wigner distribution of level spacings,
has become popular as an indicator of quantum chaos.

Apart from generic states, a subset of eigenstates selectively display pronounced 
enhancements of probability density, effectively localizing in configuration space.
Depending on their physical origin, they could be scarred or, more generally,
localized states. Scarring \cite{heller} is
a striking quantum phenomenon in chaotic systems arising from quantum interferences 
that reinforce in the vicinity of unstable periodic orbits
and survive deep in the semiclassical regime \cite{vergini}. On the other hand, the
localized states such as bouncing ball modes in billiards \cite{bill1} are induced by
classical dynamical structures.
In this paper, localized states denote any localized density enhancements
irrespective of the underlying physical mechanism that creates it. In particular,
RMT does not distinguish them based on their physical origins.
Remarkably, the class of localized states modify
the spectral statistics by inducing deviations from level repulsion property
of chaotic systems \cite{robnik1}.

Localized states were discovered in stadium billiards in 1979 \cite{sbscar} and since then 
experimentally observed in a variety of chaotic systems including deformed microcavity 
lasers \cite{mcav,jan}, quantum well with chaotic electron dynamics \cite{wilk} and hydrogen atom
in strong external fields \cite{efield}.
Recently, scarring localization was also reported in Dirac Fermions \cite{dscar},
strongly doped quantum wells \cite{scar1},
driven spin-orbit coupled cold atomic gases \cite{soatom}, a chaotic open quantum system \cite{oqs}
and in an isomerizing chemical reaction \cite{isomer}. Further, localized modes appear in
spectral graph theory in relation to random graphs \cite{graph}.

In a semiclassical sense, the generic eigenstates are associated with chaotic orbits,
and localized states with the short time periodic orbits with time scales
much shorter than the Heisenberg time $t_H \sim \hbar/\Delta$, where $\Delta$ is the
mean level spacing. This is reflected in their spectral properties as well;
nearest neighbour level spacings, $s$, of generic states are correlated and follow Wigner distribution $P_W(s)$
whereas those of localized states are nearly uncorrelated and are closer to Poisson distribution, $P_p(s)$. Are the localized levels correlated with their neighbouring generic levels?
This has not been directly probed yet. In a spectrum containing both generic and
localized states, deviations from $P_W(s)$, often modeled through use of
Brody distribution $P_B(s)$~\cite{brody}, points to the existence of non-trivial
correlations between them. This correlation quantifies the influence of localized 
states in the spectrum and indirectly characterizes
the mixed nature of the underlying classical dynamics. This is even more useful for
many-body systems, for which classical analogues may not exist but spacing distributions
are widely used to characterise the spectral properties of metallic, 
insulating, many-body localized and thermal phases \cite{mbl,rmt-mbl}.

In the semiclassical limit as $\hbar \to 0$, the fraction of localized states tends
to vanish and the spectrum is dominated by the generic states in accordance with the
semiclassical eigenfunction hypothesis \cite{eighyp} and Schnirelman theorem \cite{sch}.
Then, phenomenological distributions such as $P_B(s)$ become insensitive to the presence 
of a small subset of localized states.
In these earlier approaches, the desired correlation signal is masked by the sheer
statistical weight of the generic states.

\section{RANDOM MATRIX MODEL}

In this work, this hurdle is overcome by considering the level spacings $s_{lg}$ 
{\sl only} between localized and its neighbouring generic states.
Using a $3\times3$ random matrix model with a single parameter, for systems
that preserve, and violate time-reversal symmetry,
the exact distribution for the ratio of consecutive spacings \cite{huse,atas} is obtained.
This provides a robust characterisation of the correlation between
localized and generic states even as $\hbar \to 0$.
The analytical results are compared with the random matrix simulations as well as from
two variants of stadium billiards \cite{stadbill}, coupled quartic oscillator \cite{btu1993} and 
levels of Samarium (Sm) atom \cite{smatom1}. We note that the ratio of spacing
is a suitable statistic in this context since it does not require unfolding the spectrum.

\begin{figure}
\includegraphics*[width=3.3in]{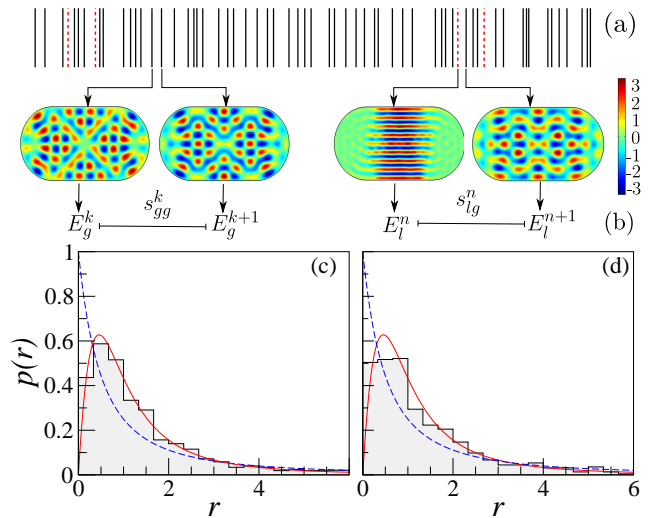}
\includegraphics*[width=3.3in]{Fig1b.eps}
\caption{(Color online) (a) Energy levels of stadium billiard. Scarred levels
are marked in dashed (red) lines. A g-g type and g-l type spacing is shown.
(b) Two consecutive generic eigenstates (state numbers 200 and 201), and
two consecutive states (245 and 246) (a generic state next to a localized state).
(c) distribution of spacing ratios for g-g type spacings, (d) distribution
of spacings for g-l type spacings. The red (solid) and blue (broken) lines
are the standard results for $p_W(r)$ and $p_{poisson}(r)$ respectively.}
\label{fig1}
\end{figure}

The main motivation behind the random matrix model can be inferred from Fig. \ref{fig1}.
A short sequence of energy levels of stadium billiards is displayed in Fig. \ref{fig1}(a) with
localized states indicated by dashed lines. In Fig. \ref{fig1}(b) two pairs of
consecutive eigenstates $|\Psi(x,y)|^2$ are shown; (i) consecutive generic states and we 
call the corresponding level spacing $s_{gg}$ to be of g-g type and (ii) localized and its
nearest neighbour generic state with spacing $s_{gl}$ of g-l type.
Such sub-sequences of levels are
commonly encountered in quantum chaotic systems with mixed classical
phase space, as well as in atomic and nuclear spectra \cite{qchatphy, qchnuphy}.
Consider a chaotic quantum system whose Hamiltonian operator is $\widehat{H}$
and its energy spectrum is $E_i$, where $i=1,2,...$ denotes
the state number. The usual approach is to analyse all the level spacings in the spectrum.
In contrast, in this work, we focus on the spacings $s_{gl}$ between generic and localized states
(Fig. \ref{fig1}(b)) defined as follows. From a sequence of consecutive energy levels $E_{k-1}<E_{k}<E_{k+1}$, where one of them corresponds to a localized
state, two spacings, $s_{k}=E_k-E_{k-1}$ and $s_{k+1}=E_{k+1}-E_{k}$, and hence one spacing ratio 
$r_k=s_k/s_{k+1}$ may be obtained, where at least one of the spacings is of the g-l type. Figure \ref{fig1}(c) shows the distribution of 
spacing ratio ${p}(r)$ obtained using only the generic levels (g-g type spacings) and in Fig. \ref{fig1}(d) for
spacings involving a localized state (g-l type). For g-g type spacings,
agreement with Wigner-type surmise
\begin{equation}
 p_W(r) = \frac{1}{Z_\beta}\frac{(r+r^2)^\beta}{(1+r+r^2)^{1+(3/2)\beta}}
\end{equation}
(with $\beta=1$ and $Z_\beta=\frac{8}{27}$)
is clearly evident whereas the g-l type spacings show marked, though weak, deviation from ${p}_W(r)$.
Hence, the Hilbert space around a localized eigenstate can be locally modeled 
as a $3 \times 3$ Hamiltonian matrix.

Thus, we are led to consider an ensemble of $3 \times 3$ real-symmetric (for $\beta=1$) or 
complex-Hermitian (for $\beta=2$) random matrices $H$ from the probability measure
\begin{equation}
\mathcal{P}(H)~d[H] \propto \exp\left(-\frac{\beta}{2}\tr\Sigma^{-2}H^2\right)~d[H].
\label{matrix_model}
\end{equation}
where `tr' represents trace, and $d[H]$ represents the product of 
differentials of all the independent parameters in the matrix elements, and 
$\Sigma=\text{diag}\left(1,1,\sqrt{\frac{k^2}{2-k^2}}\right)$. Within the framework
of Eq. \ref{matrix_model}, the random matrix is of the form
 \begin{equation}
 \mathbf{R_3} = \begin{bmatrix}
 H_{11} & H_{12} & H_{13}\\
 H_{12} & H_{22} & H_{23}\\
 H_{13} & H_{23} & H_{33}\\
 \end{bmatrix}
\label{R3}
 \end{equation}
where, $0\le k^2<2$, and $k$ is a parameter that represents the strength of 
coupling between a $2 \times 2$ 
(Gaussian Orthogonal Ensemble(GOE) or Gaussian Unitary Ensemble(GUE)) block and a $1 \times 1$ block representing a localized state.
For systems with time-reversal symmetry (TRS), i.e., $\beta=1$, the matrix elements 
are drawn from independent Gaussian distributions with mean zero and variances given as
\begin{align}
\langle H_{11}^2 \rangle =\langle H_{22}^2 \rangle =1, ~~\langle H_{33}^2 \rangle =\left(\frac{k^2}{2-k^2}\right), \nonumber \\
\langle H_{12}^2 \rangle =\frac{1}{2}, ~~\langle H_{13}^2 \rangle =\langle H_{23}^2 \rangle =\frac{k^2}{2}.
\label{matelepar}
\end{align}

For $\beta=2$, corresponding to the broken time-reversal symmetry (TRSB) case, the matrix elements are Gaussian distributed with
mean zero (real and complex for diagonal and off-diagonal), and variances given as
\begin{gather}
\begin{align}
& \expval{H_{11}^2}=\expval{H_{22}^2}=\frac{1}{2}, ~~~\expval{\Re(H_{12})^2}=\expval{\Im(H_{12})^2}=\frac{1}{4}, \nonumber
\end{align} \\
\expval{\Re(H_{13})^2}=\expval{\Im(H_{13})^2}= \frac{k^2}{4}, \nonumber \\
\expval{\Re(H_{23})^2}=\expval{\Im(H_{23})^2}=\frac{k^2}{4}, \nonumber \\
\expval{H_{33}^2}=\frac{1}{2}\left(\frac{k^2}{2-k^2}\right).
\label{b2R3}
\end{gather}

Physically, $0\le k \le 1$ indicates the strength of correlation between localized and generic states. Thus, $k <<1$ implies strong localization effects and might require semiclassical methods 
to understand its physical mechanism. On the other hand, $k \approx 1$ implies negligible localization and RMT framework would be a suitable model.
As $k\rightarrow 1$, the 3$\times$3 matrix tends to that of standard Gaussian ensembles.
In physical systems the localized and generic states are generally weakly coupled and we
anticipate the coupling strength to be weak, i.e., $k \ll 1$.
Hence, this weak coupling limit is the main regime of interest in this work.
In this limit, $H$ becomes the direct sum of a 2$\times$2 GOE or GUE matrix 
(for $\beta=1,2$, respectively) and $0$, the latter being also one of the eigenvalues and it notionally corresponds to the localized state.

\section{Distribution of spacing ratios}

The main result of the paper, namely, the distribution $p(r)$ of spacing ratios $r$
for the random matrix model defined in Eqs. (\ref{matrix_model}-\ref{b2R3}), is obtained 
analytically in this section for the cases of $\beta=1$ and $\beta=2$.

\subsection{$\beta=1$ case}

To derive an expression for $p(r)$, firstly the joint probability density of the
eigenvalues $\{\lambda\} (\equiv \lambda_1,\lambda_2,\lambda_3$) of the matrix model in 
Eq. \eqref{matrix_model} is obtained as
\begin{equation}
P(k;\{\lambda\}) \propto | \Delta(\{\lambda\})| \int_{O_3}d\mu(O)~e^{-\mbox{tr}\Sigma^{-2}O^T\Lambda^2 O} ,
\end{equation}
where $\Lambda=\mbox{diag}(\lambda_1,\lambda_2,\lambda_3)$, $\Delta(\{\lambda\})=|(\lambda_2-\lambda_1)(\lambda_3-\lambda_1)(\lambda_3-\lambda_2)|$ 
is the Vandermonde determinant and $d\mu(O)$ represents the Haar-measure over group $O_3$ of 
$3\times 3$ orthogonal matrices with $T$ being the transpose. 
 
To calculate the ratio of consecutive spacings $r$ we order them as $-\infty<\lambda_1\leq \lambda_3$, 
$-\infty< \lambda_3<\infty$, and $\lambda_3\leq \lambda_2<\infty$. Then, 
$r=(\lambda_2-\lambda_3)/(\lambda_3-\lambda_1)$. Moreover, the 
joint probability density for the ordered eigenvalues is given by 
$\widetilde{P}(k;\lambda_1,\lambda_2,\lambda)=3! P(k;\lambda_1,\lambda_2,\lambda)$, 
where the intermediate eigenvalue $\lambda_3=\lambda$.

Introducing $x=\lambda-\lambda_1$ and after some calculations whose details are in Appendix A, the distribution of $r$ can be obtained as,
\begin{eqnarray}
\label{pkr}
&&p(k;r)=\frac{\sqrt{2-k^2}}{\pi k^3}r(r+1) \int_{-\infty}^\infty d\lambda \int_0^\infty dx \int_0^{\frac{\pi}{4}} d\phi x^4 \cos \phi \nonumber \\
&&e^{-\frac{(2+k^2)}{2k^2}\lambda^2+\frac{1}{2}\big[ (\frac{1}{k^2}-1)\cos 2\phi+1\big]\big[2\lambda^2-(\lambda-x)^2-(\lambda+r x)^2\big]} \nonumber \\
&&\times ~ I_0\bigg(\frac{x}{2}\Big(\frac{1}{k^2}-1\Big)(r+1)[2\lambda+(r-1)x]\cos 2\phi\bigg).
\end{eqnarray}
As $k \to 1$, the correct GOE result $p(1;r)=\frac{27}{8}\frac{r^2+r}{\left(r^2+r+1\right)^{5/2}}$ 
is recovered as originally obtained in Ref. \cite{atas}. The limiting case $k \to 0$, relevant 
for the localized states of chaotic systems, is difficult to obtain using Eq. \eqref{pkr}. 
However, starting from the joint probability density for $k=0$ case, it is directly obtained as
\begin{eqnarray}
p(0;r) & = & \frac{1}{2 \sqrt{2}}\Bigg[ \frac{(r+1)}{\left(r^2+1\right)^{3/2}}+ \nonumber \\
& & \frac{1}{(2 r (r+1)+1)^{3/2}}+\frac{r}{(r (r+2)+2)^{3/2}} \Bigg].
\label{pr-goe}
\end{eqnarray}
In particular, note that $p(0;r)$ is different from $p_{poisson}(r) = \frac{1}{(1+r)^2}$ obtained for 
the case of uncorrelated levels with Poisson spacing distribution \cite{huse}.

\subsection{$\beta=2$ case}

For $\beta=2$, the joint probability density of (unordered) eigenvalues turns out to be
\begin{equation}
P(k;\lambda_1,\lambda_2,\lambda_3)\propto \Delta^2(\{\lambda\})\int_{\mathcal{U}_3} dU \exp\left(-\Sigma^{-2}U\Lambda^2 U^\dag\right).
\end{equation}
In this case, the unitary group integral can be performed using the Harish-Chandra-Itzykson-Zuber formula~\cite{hc1957,iz1980},
\begin{align}
\int_{\mathcal{U}_N} \,dU \exp\left(-s\tr XU Y U^\dag\right)=\prod_{m=1}^{N-1} m!\cdot \,\big(-s\big)^{-N(N-1)/2}\, \nonumber \\
 \times\frac{\det\big[\exp(-s\, x_j y_k)\big]_{j,k=1,...,N}}{\Delta(\{x\})\Delta(\{y\}) }.
\end{align}
Here, $dU$ is the Haar measure on unitary group $\mathcal{U}_N$, and $X=\text{diag}(x_1,...,x_N)$, $Y=\text{diag}(y_1,...,y_N)$.

After some calculations whose details may be found in Appendix A, the distribution of the ratio of spacings can be obtained as
\begin{eqnarray}
p(k;r)=\frac{\sqrt{2-k^2}}{4\pi k(1-k^2)^2}r(r+1)\sum_{j=1}^3 \Bigg[ \frac{b_j(5a_j^2+2b_j^2)}{a_j^4(a_j^2+b_j^2)^2}+ \nonumber \\
\frac{3}{(a_j^2+b_j^2)^{5/2}}\sinh^{-1}\Big(\frac{b_j}{a_j}\Big)-\frac{c_j(5a_j^2+2c_j^2)}{a_j^4(a_j^2+c_j^2)^2}- \nonumber \\ \label{gue}
\frac{3}{(a_j^2+c_j^2)^{5/2}}\sinh^{-1}\Big(\frac{c_j}{a_j}\Big)\Bigg]. ~~~
\label{gue-res}
\end{eqnarray}
The forms of $a_j, b_j$ and $c_j$ for $j=1,2,3$ (which are functions of $k$ and $r$) 
are rather unwieldy and is shown in Appendix A.
In the limit $k \to 0$, the exact result is obtained as
\begin{eqnarray}
p(0;r) & = & \frac{1}{\pi} \Bigg[ \frac{r^2}{(r (r+2)+2)^2}+  \nonumber \\
 & & \frac{r(r+2)+1}{ \left( r^2+1\right)^2}+\frac{1}{(2 r (r+1)+1)^2} \Bigg].
\label{pr-gue}
\end{eqnarray}
As anticipated, when $k \to 1$, the distribution in Eq. \eqref{gue-res} coincides 
with the GUE result, obtained in Ref. \cite{atas}, namely, $p(1;r)=\frac{81 \sqrt{3}}{4 \pi}\frac{ \left(r^2+r\right)^2}{ \left(r^2+r+1\right)^4}$.

\section{Numerical simulation results}

\subsection{Random Matrix Model}

\begin{figure}
\includegraphics*[width=3.6in,angle=0]{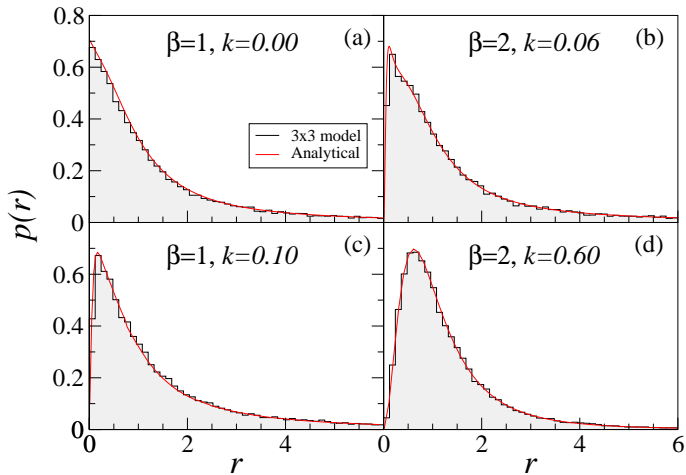}
\caption{(Color online) Spacing ratio distribution, for g-l type spacings, obtained 
from random matrix simulations of $3 \times 3$ 
random matrices (histogram) compared with analytical $p(r)$ (red line). 
See text for how g-l type spacings are identified in simulations.
The $\beta=1$ and $\beta=2$ cases are shown.}
\label{fig2}
\end{figure}

In Fig. \ref{fig2}, the analytically obtained $p(r)$ in 
Eqs. \eqref{pkr} and \eqref{gue-res} is compared with the results obtained by simulating
an ensemble of $3 \times 3$ random matrices $\mathbf{R}_3$ following the prescription 
in Eqs. \eqref{matrix_model} and \eqref{matelepar}. The numerical simulations are 
performed by generating matrix
elements with prescribed mean and variances. If $k=0$, $\mathbf{R}_3$ is block
diagonal; one $2 \times 2$ block with eigenvalues $\lambda_1, \lambda_2$ and a $1 \times 1$ block 
with eigenvalue $\lambda_3$. The eigenvalue corresponding to a localized state
is identified using the information entropy of an eigenstate~\cite{infent}. The eigenvector corresponding 
to $\lambda_i, (i=1,2,3),$ is $(a_{i,1}, a_{i,2}, a_{i,3})$. The corresponding information entropy is
$S_i = - \sum_j |a_{i,j}|^2 \ln |a_{i,j}|^2$. For $k=0$, the eigenvalue whose eigenvector is 
$(0,0,1)$ is far from a generic state and hence can be called `localized' eigenvalue for our purposes.
In this case, $S=0.0$. As $k \to 1$, localized states typically disappear from the spectrum.
If $\lambda$ is the eigenvalue of the localized state identified using information 
entropy, then the spacing ratio is calculated as either $r = (\lambda - \lambda_2)/(\lambda_2 - \lambda_1),
r = (\lambda_3 - \lambda)/(\lambda - \lambda_1)$, or $r = (\lambda_3 - \lambda_2)/(\lambda_2 - \lambda)$, depending on whether the localized 
state corresponds to $\lambda_3$, $\lambda_2$ or $\lambda_1$ respectively.
In Fig. \ref{fig2}(b,d), simulated histograms of $p(r)$ for $\beta=2$ is shown and
displays an excellent agreement with Eq. \eqref{gue-res}.

\subsection{Applications to Physical Systems}

\begin{figure}[t]
\includegraphics*[width=3.3in,angle=0]{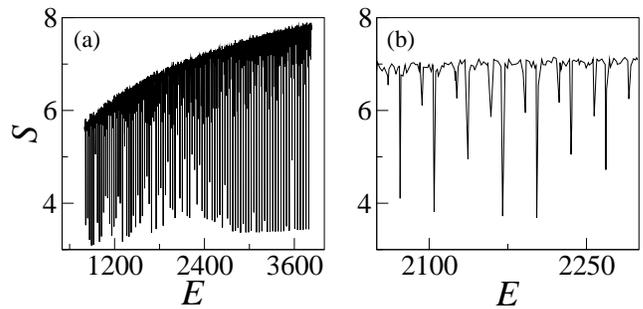}
\caption{(a) Information entropy(S) as a function of energy(E) for the coupled quartic oscillator system at $\alpha$=90. The eigenstates having
magnitude of information entropy $\lesssim 5.5$ can be identified as localized states. 
For the bulk of chaotic states that form the envelope, value of $S$ is consistent with 
the random matrix average for the information entropy (not shown here). 
(b) Enlarged view of a portion of (a), consisting of
175 states, out of which 4 may be considered to be localized.}
\label{enent}
\end{figure}

The spacing ratio distribution for g-l type spacings is
obtained for Hamiltonian systems whose classical limit is chaotic and hence
their spacings are Wigner distributed, $P_W(s)$.
The systems chosen for illustration are (i) the coupled quartic oscillator,
(ii) computed levels of Sm atom, and (iii) stadium billiards ($\beta=1$ and $\beta=2$ variants).
All of them contain localized eigenstates in their spectrum.
The computed distribution for g-l type spacing ratios
agrees with the analytical results and in this $k$ is treated as a fitting parameter.

\subsubsection{Coupled Quartic Oscillator}

In Fig. \ref{fig3}(a), the results are displayed for the coupled quartic oscillator, 
a well studied model of quantum chaos \cite{atkins}. The Hamiltonian for this system is 
\begin{equation}
 H=p_x^2 + p_y^2 + x^4 + y^4 + \alpha x^2 y^2,
\end{equation}
where $\alpha$ is the chaos parameter. 
The system becomes increasingly more chaotic as $\alpha \to \infty$.
In this work, $\alpha$ is chosen to be equal to $90$ such that the classical phase space is largely
chaotic, with small regular regions due to the presence of a series of periodic orbits studied in detail in
Ref \cite{qoloc}. This is manifested in the corresponding quantum system
as localized eigenstates, which are identified using information entropy. 
This is illustrated in Fig \ref{enent}
using the information entropy calculated for about 1800 states of the quartic oscillator at $\alpha=90$. In Fig. \ref{fig3}(a), the computed distribution of 
g-l type spacings for coupled quartic oscillator displays a good agreement with the 
analytical result shown in Eq. \ref{pkr} with $k \approx 0.2$.
As an independent verification, $k$ can be extracted from the variance of the
off-diagonal matrix elements (of Hamiltonian operator) locally around every localized state.
For quartic oscillator, both estimates agree with each other to within 30\% error.

\subsubsection{Sm atom}

\begin{figure}[t]
\includegraphics*[width=3.4in,angle=0]{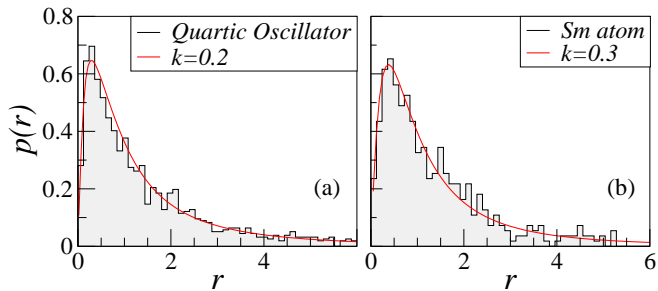}
\caption{(Color online) Spacing ratio distribution, for g-l type spacings, obtained 
from systems whose classical limit is chaotic.
Histograms are obtained from spectrum computed for (a) quartic oscillator and 
(b) levels of Sm from ab-initio calculations. The solid red) line
is the fit obtained using the analytical relation in Eq. \eqref{pkr}.}
\label{fig3}
\end{figure}

The application of random matrix theory in quantum chaos was first motivated by the study of spectral fluctuations in complex nuclei \cite{wigner}. 
Lanthanide atoms, like Samarium (Sm) have been studied in this context, using the multi-configuration Dirac-Fock method to compute their spectra
and identify localized states. Localization in this context, is known to occur due to strong Coulomb mixing between configuration state functions
having similar occupancy numbers of their subshells \cite{smatomloc}. Figure \ref{fig3}(b) shows $p(r)$ for g-l type spacings in the computed levels of
a lanthanide atom, namely, Samarium (Sm). The energy levels of Sm exhibits complex configuration and
mixing and were computed using {\sc GRASP} code \cite{grasp} The computed 
histogram of g-l type spacing ratios agree with $p(r)$ for $k \approx 0.3$.

\subsubsection{Quantum Billiards with and without time-reversal symmetry}

\begin{figure}[t]
\includegraphics*[width=3.4in,angle=0]{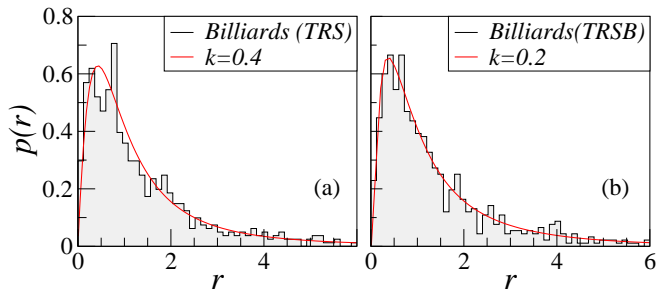}
\caption{(Color online) Spacing ratio distribution, for g-l type spacings, obtained 
from systems whose classical limit is chaotic. Histograms are obtained from spectrum 
computed for stadium billiards, with time-reversal symmetry (a) preserved ($\beta=1$)
and (b) broken ($\beta=2$). The solid (red) line is the fit obtained using the 
analytical relation in Eqs. \eqref{pkr} and \eqref{gue-res} for (a) and (b) respectively.}
\label{fig4}
\end{figure}

The ratio distribution $P(r)$ for quarter stadium billiards is shown
in Fig. \ref{fig4}(a,b), respectively, for TRS and TRSB cases. The eigenvalues for a 
closed quantum billiard system may be obtained by solving the Helmholtz equation
\begin{equation}
 [\vec{\nabla}^2+\mu^2]\vec{\mathit{E}}=0
\end{equation}
with appropriate boundary conditions, where $\vec{\mathit{E}}$ is the electric field and $\mu$ the wavenumber.
Localized and scarred states in billiards had been extensively investigated earlier \cite{scarth,scarexp}
and deviations from $P_W(s)$ is attributed to such states, of which bouncing ball modes 
form a prominent class \cite{stadbillbb}. In Fig. \ref{fig4}(a), the simulated histogram of g-l type spacing for billiards agrees with the analytical result in Eq. \eqref{pkr}
for $k \approx 0.4$. The TRSB case, corresponding to
GUE results, is illustrated using a stadium billiard with a magnetized ferrite
strip placed perpendicular to the two horizontal boundaries of the stadium. The
TRS is broken by the phase shifts induced in the incident
electromagnetic waves upon reflection from boundaries in the presence of static
magnetic field. This has been experimentally realized in microwave cavities \cite{stodietz,so}, 
and has been simulated here with parameters taken from \cite{so}, modifying only the shape 
to a quarter stadium, in order to realize the bouncing ball modes, similar to the TRS case. 

In order to compute the g-l type spacings, firstly the eigen spectra of this system was
computed using a commercial finite element method software, COMSOL Multiphysics \cite{comsol}.
Then, using the computed eigenvalues and normalized eigenvectors, i.e. the magnitude of the 
electric field $E$, the information entropy for each state is calculated using the relation
$S_i = - \sum_j |E_{i,j}|^2 \ln |E_{i,j}|^2$, where $j$ is the index for discretized position
space. The localized states may be differentiated from 
the bulk of the chaotic states since the former have a significantly smaller magnitude of 
information entropy compared to the latter. Among all the localized states, bouncing ball modes are observed to be the 
most strongly localized, and the hence the information entropy corresponding to these states 
has the least magnitude. Using the information entropy, the localized states are picked from the
spectrum (as illustrated in Fig \ref{enent}(b)), and the required ratio distribution is determined.
The result for $p(r)$ is shown as histogram in Fig. \ref{fig4}(b) and it agrees with the analytical result
(Eq. \eqref{gue-res}) with $k \approx 0.2$.

\section{CONCLUSION}

To summarize, typical spectrum of a chaotic quantum system has
generic and localized eigenstates occurring as neighbours. Physically, they
represent two distinct limiting behaviours. The former is modeled by random matrix
assumptions and the latter deviates strongly from RMT based models.
In this work, it is demonstrated that they display non-trivial
correlations, quantified by the parameter $k$, the average strength of
the Hamiltonian matrix element coupling these states. Physically, $k$ is
a measure of the strength of correlation between localized and generic states.

This is a robust characterization that remains unaffected by the semiclassical limit
in contrast to the phenomenological approach such as Brody distribution had 
often been used to model the spectral transition from Poisson to GOE type statistics. 
In such an analysis, all the levels (localized and generic) are taken into account. 
Then, in the semiclassical limit of $\hbar \to 0$ or energy $E \to \infty$, localized 
modes ultimately would become a set of measure zero and the Brody distribution would
nearly coincide with random matrix distributions. Hence, signatures of the localized states 
are masked by the large number of chaotic states.
Thus, the Brody parameter being a single number representing this transition would become 
insensitive to presence of localized modes in the semiclassical limit.
In the approach presented in this paper, since the spacings (by construction)
always involve at least one
localized mode, the estimated value of $k$ remains unaffected by the semiclassical limit.

Hence, the parameter $k$ can be thought of as representing the coupling between a
generic and a localized eigenstate and is sensitive even as $\hbar \to 0$.
In this work, by considering a $3 \times 3$ random matrix model, an exact result 
is obtained for the distribution of the 
spacing ratio $s_{lg}$ between a generic and a localized state.
The analytical results are in good agreement with numerically computed spectra obtained
from chaotic quantum systems such as billiards, coupled oscillator and atomic spectra.
In practice, it is not necessary to compute the localized states to estimate the
value of the parameter $k$. For most physical systems which display localized states,
adiabatic methods can estimate the energies of localized states without computing the
eigenvectors and information entropy. Such results exist for quartic 
oscillator \cite{qoloc,mss-adia} and stadium billiards \cite{bill-adia}.
Quantum stadium billiard had been experimentally realized
and hence the results presented here can be experimentally verified as well.

\onecolumngrid
\appendix
    
\section{Generalized Gaussian ensemble and ratio of consecutive level spacings}

In order to derive the distribution of ratio of consecutive eigenvalue spacings, 
we need the joint probability density of the eigenvalues 
of the matrix model defined by Eq. \eqref{matrix_model}. 
The cases of $\beta=1$ and $\beta=2$ are dealt with separately below.

\subsection{$\beta=1$ ($2\times 2$ GOE $ \oplus$ Localized $\rightarrow$ $3\times 3$ GOE)}

The joint probability density of eigenvalues in this case follows as
\begin{equation}
\label{beta1jpd}
P(k;\lambda_1,\lambda_2,\lambda_3)\propto |\Delta(\{\lambda\})|\int_{\mathcal{O}_3} dO \exp\left(-\frac{1}{2}\Sigma^{-2}O\Lambda^2 O^T\right),
\end{equation}
where the integral is over the group of $3\times 3$ orthogonal matrices with $dO$ representing the corresponding Haar measure. 
Also, $\Delta(\{z\})=\prod_{j>k}(z_j-z_k)=\det[z_k^{j-1}]_{j,k=1,...,N}$ is the Vandermonde determinant. For the unitary group, the celebrated
Harischandra-Itzykson-Zuber formula~\cite{hc1957,iz1980} gives the result for this integral. Here we do not have such a result because $\Sigma^{-2}$
and $\Lambda^2$ do not lie in the Cartan subalgebra corresponding to the orthogonal group. 
Nevertheless, for the $3$-dimensional case, it is possible to make progress using the recursive approach suggested by Guhr and Kohler~\cite{gko}. 
Then, we have
\begin{gather}
\label{O3int}
\nonumber
\int_{\mathcal{O}_3} \,dO \exp\left(-s\tr XO Y O^T\right)=\frac{1}{2\pi} \exp(-s(x_1+x_2+x_3)y_3) \nonumber \\
\times \int_{x_1}^{x_2}dx'_1 \int_{x_2}^{x_3}dx'_2\frac{(x'_2-x'_1)}{\bigg(-\displaystyle\prod_{{j=1,2,3}\atop{k=1,2}}(x_j-x'_k)\bigg)^{1/2}} \nonumber \\
\times \exp\left(-s(x'_1+x'_2) \left(\frac{y_1+y_2}{2}-y_3 \right) \right) \nonumber \\
 \times I_0\left(\frac{s\,(x'_1-x'_2)(y_1-y_2)}{2}\right),
\end{gather}
where $I_0(z)$ is the modified Bessel function of the first kind and zeroth order \cite{bessel}.
This result cannot be used directly for Eq.~\eqref{beta1jpd} since $\Sigma^{-2}$ has two identical entries. 
We need to consider the limit $x_1,x_2\rightarrow 1$. For this let us set $x_1=1-\epsilon$ and $x_2=1$ and take the limit $\epsilon\rightarrow 0$.
The crucial part in the above expression is
\begin{align*}
&\lim_{\epsilon\rightarrow 0}\int_{1-\epsilon}^1 dx'_1 \frac{1}{\big[-(1-\epsilon-x'_1)(1-x'_1)\big]^{1/2}}\\
&=\lim_{\epsilon\rightarrow 0}\int_{1-\epsilon}^1 dx'_1 \frac{1}{\big[-(1-\epsilon-x'_1)(1-x'_1)\big]^{1/2}}\\
&=2\lim_{\epsilon\rightarrow 0}\int_0^{\pi/2} d\theta\,,~~~~~~~\text{where } \epsilon\sin^2\theta=1-x'_1\\
&= \pi.
\end{align*}
The other occurrences of $x'_1$ can be taken as 1.
Now, using Eq.~\eqref{beta1jpd}, substituting $s=1/2, x_3=(2-k^2)/k^2, y_1=\lambda_1^2, y_2=\lambda_2^2, y_3=\lambda_3^2$, calling $x'_2=u$, and fixing the normalization the joint probability density of eigenvalues is obtained as:
 \begin{align}
\label{PLO}
\nonumber
\nonumber
P(k;\lambda_1,\lambda_2,\lambda_3)=\frac{\sqrt{2-k^2}}{24\pi k^2\sqrt{1-k^2}} |(\lambda_2-\lambda_1)(\lambda_3-\lambda_1)(\lambda_3-\lambda_1)|e^{-\big(\frac{2+k^2}{2k^2}\big)\lambda_3^2}\\
\times \int_1^{2/k^2-1}\!\!du\, \frac{1}{\sqrt{2/k^2-1-u}}e^{-\frac{(u+1)}{4}(\lambda_1^2+\lambda_2^2-2\lambda_3^2)}I_0\left(\frac{(u-1)}{4}(\lambda_1^2-\lambda_2^2)\right).
\end{align}

For calculating the ratio, the eigenvalues are restricted to the region defined by $-\infty<\lambda_1\leq \lambda_3$, $-\infty< \lambda_3<\infty$, $\lambda_3\leq \lambda_2<\infty$. 
The joint probability density of these ordered eigenvalues is then given by
\begin{align}
\widetilde{P}(k;\lambda_1,\lambda_2,\lambda_3)=3! P(k;\lambda_1,\lambda_2,\lambda_3).
\end{align}
The probability density function of the ratio of consecutive spacings $r=(\lambda_2-\lambda_3)/(\lambda_3-\lambda_1)$ can then be found as
\begin{align}
p(k;r)=\int_{-\infty}^\infty d\lambda_3 \int_{-\infty}^{\lambda_3} d\lambda_1  &\int_{\lambda_3}^\infty d\lambda_2\,\delta\left(r-\frac{\lambda_2-\lambda_3}{\lambda_3-\lambda_1}\right) \nonumber \\
&\times\widetilde{P}(k;\lambda_1,\lambda_2,\lambda_3).
\end{align}
Let us call $\lambda_3=\lambda$ and define $x=\lambda-\lambda_1$ and $y=\lambda_2-\lambda$, then the above integral, in terms of these new variables, becomes
\begin{align}
p(k;r)=\int_{-\infty}^\infty d\lambda \int_0^\infty &dx\int_0^\infty dy\,\delta\left(r-\frac{y}{x}\right) \nonumber \\
&\times\widetilde{P}(k;\lambda-x,\lambda+y,\lambda).
\end{align}
The delta function integral over $y$ can be trivially done to yield
\begin{align}\label{common}
p(k;r)=\int_{-\infty}^\infty d\lambda \int_0^\infty dx\,\,x \widetilde{P}(k;\lambda-x,\lambda+r x,\lambda).
\end{align}
Using Eq.~\eqref{PLO} in this, we obtain
 \begin{align}
\nonumber
p(k;r)=\frac{\sqrt{2-k^2}}{4\pi k^2\sqrt{1-k^2}}r(r+1)& \int_{-\infty}^\infty d\lambda \int_0^\infty dx \int_1^{2/k^2-1} \!\!du \,x^4 e^{-\frac{(2+k^2)}{2k^2}\lambda^2+\frac{1}{4}(u+1)[2\lambda^2-(\lambda-x)^2-(\lambda+r x)^2]}\\
&\times \Big(\frac{2-k^2}{k^2}-u\Big)^{-1/2}I_0\Big(\frac{1}{4}x(u-1)(r+1)[2\lambda+(r-1)x]\Big).
\end{align}

It is found that the substitution $u=1+2(1/k^2-1)\cos 2\phi$ leads to an expression which is comparatively more stable for numerical computation purposes :
\begin{align}
\nonumber
p(k;r)=\frac{\sqrt{2-k^2}}{\pi k^3}r(r+1)& \int_{-\infty}^\infty d\lambda \int_0^\infty dx \int_0^{\pi/4} \!\!d\phi \,x^4 \cos \phi \,e^{-\frac{(2+k^2)}{2k^2}\lambda^2+\frac{1}{2}\big[ (\frac{1}{k^2}-1)\cos 2\phi+1\big]\big[2\lambda^2-(\lambda-x)^2-(\lambda+r x)^2\big]}\\
&\times I_0\bigg(\frac{1}{2}x\Big(\frac{1}{k^2}-1\Big)(r+1)[2\lambda+(r-1)x]\cos 2\phi\bigg) .
\end{align}

\subsection{$\beta=2$ ($2\times 2$ GUE $ \oplus$ Localized $\rightarrow$ $3\times 3$ GUE)}
The joint probability density of (unordered) eigenvalue in this case follows as
\begin{equation}
P(k;\lambda_1,\lambda_2,\lambda_3)\propto \Delta^2(\{\lambda\})\int_{\mathcal{U}_3} dU \exp\left(-\Sigma^{-2}U\Lambda^2 U^\dag\right).
\end{equation}
In this case, the unitary group integral can be performed using the Harish-Chandra-Itzykson-Zuber formula~\cite{hc1957,iz1980},
\begin{align}
\int_{\mathcal{U}_N} \,dU \exp\left(-s\tr XU Y U^\dag\right)=\prod_{m=1}^{N-1} m!\cdot \,\big(-s\big)^{-N(N-1)/2}\, \nonumber \\
\times\frac{\det\big[\exp(-s\, x_j y_k)\big]_{j,k=1,...,N}}{\Delta(\{x\})\Delta(\{y\}) }.
\end{align}
Here, $dU$ is the Haar measure on unitary group $\mathcal{U}_N$, and $X=\text{diag}(x_1,...,x_N)$, $Y=\text{diag}(y_1,...,y_N)$. If there is some multiplicity in the entries of $X$ or $Y$, 
then we must use the above formula using proper limits. This is the case here, as $\Sigma^{-2}$ has two identical entries, viz. 1. We find


\begin{align}
\nonumber
P(k;\lambda_1,\lambda_2,\lambda_3)&\propto (\lambda_2-\lambda_1)^2(\lambda_3-\lambda_1)^2(\lambda_3-\lambda_2)^2\frac{\det\begin{bmatrix}e^{-\lambda_1^2} & -\lambda_1^2e^{- \lambda_1^2} & e^{-(\frac{2-k^2}{k^2})\lambda_1^2}\\ 
e^{-\lambda_2^2} & -\lambda_2^2 e^{- \lambda_2^2} & e^{-(\frac{2-k^2}{k^2})\lambda_2^2}\\
e^{-\lambda_3^2} & -\lambda_3^2 e^{- \lambda_3^2} & e^{-(\frac{2-k^2}{k^2})\lambda_3^2}
\end{bmatrix}}{(\lambda_2^2-\lambda_1^2)(\lambda_3^2-\lambda_1^2)(\lambda_3^2-\lambda_2^2)\,\det\begin{bmatrix}1 & 0 & 1 \\ 1 & 1 & (\frac{2-k^2}{k^2}) \\ 1 & 2 & (\frac{2-k^2}{k^2})^2\end{bmatrix}}\\
&\propto \frac{(\lambda_1-\lambda_2)(\lambda_2-\lambda_3)(\lambda_3-\lambda_1)}{(\lambda_1+\lambda_2)(\lambda_2+\lambda_3)(\lambda_3+\lambda_1)}\det\begin{bmatrix}
e^{-\lambda_1^2} & \lambda_1^2e^{- \lambda_1^2} & e^{-(\frac{2-k^2}{k^2})\lambda_1^2}\\ 
e^{-\lambda_2^2} & \lambda_2^2e^{- \lambda_2^2} & e^{-(\frac{2-k^2}{k^2})\lambda_2^2}\\
e^{-\lambda_3^2} & \lambda_3^2e^{- \lambda_3^2} & e^{-(\frac{2-k^2}{k^2})\lambda_3^2}
\end{bmatrix}.
\end{align}
On expanding the determinant and fixing the normalization factor, we get
 \begin{align}\label{PLU}
\nonumber
P(k;\lambda_1,\lambda_2,\lambda_3)&=-\frac{\sqrt{2-k^2}}{3\pi^{3/2}k(1-k^2)^2}\frac{(\lambda_1-\lambda_2)(\lambda_2-\lambda_3)(\lambda_3-\lambda_1)}{(\lambda_1+\lambda_2)(\lambda_2+\lambda_3)(\lambda_3+\lambda_1)} \,e^{-\lambda_1^2-\lambda_2^2-\lambda_3^2}\\
&\times\Big[e^{-2\big(\frac{1}{k^2}-1\big)\lambda_1^2}(\lambda_2^2-\lambda_3^2)+e^{-2\big(\frac{1}{k^2}-1\big)\lambda_2^2}(\lambda_3^2-\lambda_1^2)+e^{-2\big(\frac{1}{k^2}-1\big)\lambda_3^2}(\lambda_1^2-\lambda_2^2)\Big].
\end{align}
Proceeding similar to $\beta=1$ case, we have Eq.~\eqref{common}, but with $\widetilde{P}(k;\lambda_1,\lambda_2,\lambda_3)$ obtained from Eq.~\eqref{PLU}. We find that
\begin{align}
& p(k;r)=\frac{2\sqrt{2-k^2}}{\pi^{3/2}k(1-k^2)^2}r(r+1)\int_{-\infty}^\infty d\lambda_3 \nonumber \\
&\times\int_0^\infty dx\,\,[t_1(\lambda,x)+t_2(\lambda,x)+t_3(\lambda,x)],
\end{align}
where 
\begin{equation*}\label{t1}
t_1(\lambda,x)=\frac{x^5 e^{-(1-r^2+\frac{2r^2}{k^2})x^2+2(1+r-\frac{2r}{k^2})x \lambda-(1+\frac{2}{k^2})\lambda^2}}{[(r-1)x+2\lambda](r x+2\lambda)},
\end{equation*}
\begin{equation*}\label{t2}
t_2(\lambda,x)=-\frac{(r+1)\,x^5 e^{-(1+r^2)x^2+2(1-r)x \lambda-(1+\frac{2}{k^2})\lambda^2}}{(-x+2\lambda)(rx+2\lambda)},
\end{equation*}
\begin{equation*}\label{t3}
t_3(\lambda,x)=\frac{r\,x^5  e^{-(r^2+\frac{2}{k^2}-1)x^2+2(\frac{2}{k^2}-r-1)x \lambda-(1+\frac{2}{k^2})\lambda^2}}{(-x+2\lambda)[ (r-1)x+2\lambda]}.
\end{equation*}
We notice that integrals involving $t_1,t_2,t_3$ are of a similar form, as given below:

\begin{align}\label{int}
\int_{-\infty}^\infty d\lambda \int_0^\infty dx\, \frac{x^5e^{-\alpha^2 x^2+2\eta x\lambda-\gamma^2\lambda^2}}{(u x+2\lambda)(vx+2\lambda)}=\frac{\sqrt{\pi } }{8(v-u)}\left[\frac{b \left(5 a^2+2 b^2\right)}{a^4 \left(a^2+b^2\right)^2}+\frac{3 \sinh ^{-1}\left(\frac{b}{a}\right)}{ \left(a^2+b^2\right)^{5/2}}-\frac{c \left(5 a^2+2c^2\right)}{a^4 \left(a^2+c^2\right)^2}-\frac{3\sinh ^{-1}\left(\frac{c}{a}\right)}{ \left(a^2+c^2\right)^{5/2}}\right]. 
\end{align}

Here $a^2=\alpha^2-\frac{\eta^2}{\gamma^2}$, $b=\frac{\gamma}{2}\left(u+\frac{2\eta}{\gamma^2}\right)$, $c=\frac{\gamma}{2}\left(v+\frac{2\eta}{\gamma^2}\right)$. The integral converges for $\alpha^2>0,\gamma^2>0,\alpha^2\gamma^2-\eta^2>0$. Hence, we can write down a closed form result for $p(k;r)$ based on this integral. Define

 \begin{align}\label{coeff}
\nonumber
& a_1=\frac{\sqrt{2 [1+r(r+1)(2-k^2)]}}{\sqrt{2+k^2}},~~a_2=\frac{\sqrt{2\big[1+r(r+k^2)\big]}}{\sqrt{2+k^2}},~~a_3=\frac{\sqrt{2\big[2+r(r+2)-k^2(r+1)\big]}}{\sqrt{2+k^2}},\\
\nonumber
& b_1=\frac{k^2(3r+1)-2(r+1)}{2k\sqrt{2+k^2}},~~~~~b_2=\frac{2+k^2(2r-1)}{2k\sqrt{2+k^2}},~~~~b_3=\frac{2-k^2(2r+3)}{2k\sqrt{2+k^2}},\\
& c_1=\frac{k^2(3r+2)-2r}{2k\sqrt{2+k^2}},~~~~~c_2=\frac{k^2(r-2)-2r}{2k\sqrt{2+k^2}},~~~~~c_3=\frac{2(r+1)-k^2(r+3)}{2k\sqrt{2+k^2}}.
\end{align}

Then the PDF for ratio of spacings is given by
 \begin{align}
\nonumber
p(k;r)=\frac{\sqrt{2-k^2}}{4\pi k(1-k^2)^2}r(r+1)\sum_{j=1}^3 \Bigg[ \frac{b_j(5a_j^2+2b_j^2)}{a_j^4(a_j^2+b_j^2)^2}+\frac{3}{(a_j^2+b_j^2)^{5/2}}\sinh^{-1}\Big(\frac{b_j}{a_j}\Big)\\
-\frac{c_j(5a_j^2+2c_j^2)}{a_j^4(a_j^2+c_j^2)^2}-\frac{3}{(a_j^2+c_j^2)^{5/2}}\sinh^{-1}\Big(\frac{c_j}{a_j}\Big)\Bigg].
\end{align}

It should be noted that the factor $(v-u)$ in the denominator of Eq.~\eqref{int} is $1$, $r+1$, and $r$, respectively, for the integrals involving $t_1, t_2$, and $t_3$. The third one cancels the $r$ factor in the numerator of Eq.~\eqref{t3}, while the second one, when combined with $r+1$ in the numerator of Eq.~\eqref{t2}, leaves an overall negative sign. This sign has been absorbed in the definitions for $b_2$ and $c_2$ in Eq.~\eqref{coeff}, noting that $\sinh^{-1} z$ in an odd function of $z$. 

\twocolumngrid

\end{document}